\documentclass[12pt]{article}
\usepackage{graphicx}
\usepackage{latexsym}
\def\be{\begin{equation}}
\def\ee{\end{equation}}
\def\N{{\cal N}}
\def\barf#1{#1}%
\def\bea{\begin{eqnarray}}
\def\eea{\end{eqnarray}}
\begin{document}

\begin{flushright}
\today\\
arXiv:0708.4321
\end{flushright}

\begin{center}
{\Large \bf Non-Gaussianities from ekpyrotic collapse} \\
{\Large \bf with multiple fields}
\vskip 1cm

Kazuya Koyama$^{\dagger}$\footnote{ E-mail: Kazuya.Koyama@port.ac.uk},
Shuntaro Mizuno$^{\ddagger}$\footnote{ E-mail: mizuno@resceu.s.u-tokyo.ac.jp},
Filippo Vernizzi$^{*}$\footnote{E-mail: vernizzi@ictp.it},
and
David Wands$^{\dagger}$\footnote{E-mail: David.Wands@port.ac.uk }

\vskip 1cm $^{\dagger}$ Institute of Cosmology and Gravitation, Mercantile House,
University of Portsmouth, Portsmouth~PO1~2EG, United Kingdom \\
$^{\ddagger}$  Research Center for the Early Universe (RESCEU), School of
Science, University of Tokyo, 7-3-1 Hongo, Bunkyo, Tokyo~113-0033,
Japan\\
$^*$  Abdus Salam ICTP, Strada Costiera 11, 34100 Trieste, Italy
\end{center}

\begin{abstract}
We compute the non-Gaussianity of the curvature perturbation
generated by ekpyrotic collapse with multiple fields.
The transition from the multi-field scaling solution to a
single-field dominated regime converts initial isocurvature field
perturbations to an almost scale-invariant comoving curvature
perturbation.
In the specific model of two fields, $\phi_1$ and $\phi_2$, with
exponential potentials, $-V_i \exp (-c_i \phi_i)$, we calculate the
bispectrum of the resulting curvature perturbation.
%
We find that the non-Gaussianity is dominated by non-linear
evolution on super-Hubble scales and hence is of the local form.
The non-linear parameter of the curvature perturbation is given by
$f_{NL} = -5 c_j^2 /12$, where $c_j$ is the exponent of the potential
for the field which becomes sub-dominant at late times. Since
$c_j^2$ must be large, in order to generate an almost scale
invariant spectrum, the non-Gaussianity is inevitably large. 
By combining the present observational
constraints on $f_{\rm NL}$ and the scalar spectral index, the specific model studied in this paper is thus ruled out by current observational data.
\end{abstract}

\section{Introduction}

The existence of an almost scale-invariant spectrum of primordial
curvature perturbations on large scales, with an approximately
Gaussian distribution, is one of the most important observations
that any model of the early universe should explain. The
inflationary scenario offers a possible explanation in terms of
vacuum fluctuations of scalar fields during an accelerated expansion
preceding the standard hot big bang, though attempts to embed such a
scenario within a fundamental theory such as superstring/M-theory
may require some degree of fine-tuning (see \cite{Linde:2007fr} for
a recent review).

The ekpyrotic scenario \cite{Khoury:2001wf} (see also
\cite{Kallosh:2001ai,Khoury:2001iy}) is one alternative approach
where the large-scale perturbations are generated from vacuum
fluctuations during a collapse phase driven by a scalar field with a
steep, negative exponential potential. It was shown, however, that
even though the Bardeen potential acquires a scale-invariant
spectrum during ekpyrotic collapse, the comoving curvature
perturbation has a steep blue spectrum in the original ekpyrotic
scenario driven by a single scalar field \cite{Lyth:2001pf}. If the
contracting pre-big bang phase is connected to the expanding hot big
bang phase through a regular four-dimensional bounce, then we expect
the comoving curvature perturbation to remain constant for adiabatic
perturbations on large scales
\cite{Lyth:2003im,Creminelli:2004jg,Copeland:2006tn}, which means
that the growing mode of curvature perturbations in the expanding
phase also acquires a steep blue spectrum.

One way to avoid this is to consider non-adiabatic perturbations,
which require two or more fields \cite{Gordon:2000hv,Notari:2002yc}.
Recently, there has been progress in generating a scale-invariant
spectrum for curvature perturbations in the ekpyrotic scenario
with more than one field,
which we will refer to as the new ekpyrotic scenario
\cite{Lehners:2007ac,Buchbinder:2007ad,Creminelli:2007aq}. If these
fields have steep negative exponential potentials, there exists a
scaling solution where the energy densities of the fields grow at
the same rate during the collapse \cite{Finelli:2002we,Guo:2003eu}.
In this multi-field scaling solution background, the isocurvature
field perturbations have an almost scale-invariant spectrum
\cite{Finelli:2002we}, owing to a tachyonic instability in the
isocurvature field. The multi-field scaling solution in the new
ekpyrotic scenario can be shown to be an unstable saddle point in
the phase space and the stable late-time attractor is the old
ekpyrotic collapse dominated by a single field \cite{KW}.

The existence of a tachyonic instability raises questions about
initial conditions in the new ekpyrotic scenario
\cite{Buchbinder:2007tw}, but the transition from the multi-field
scaling solution to the single-field-dominated solution also
provides a mechanism to automatically convert the initial
isocurvature field perturbations about the multi-field scaling
solution into comoving curvature perturbations about the late-time
attractor \cite{KMW}. In this case, the final amplitude of the
comoving curvature perturbation is determined by the Hubble scale at
the transition and if this parameter and the initial conditions are
set appropriately, the prediction for the primordial curvature
perturbations from this scenario
is compatible with an almost scale-invariant spectrum (see
\cite{Lehners:2007ac,Buchbinder:2007ad,Creminelli:2007aq,Tolley:2007nq}
for other mechanisms to convert the initial scale-invariant
isocurvature perturbations into curvature perturbations).

Recently, the non-Gaussianity of the distribution of primordial
curvature perturbations in the inflationary scenario has been
extensively studied by many authors (see e.g.~\cite{Bartolo:2004if}
for a review). Measurements of the non-Gaussianity already provide
important constraints on specific models of the early universe and
such measurements will continue to improve in the near future, for
instance with the Planck satellite \cite{Planck}. Since the
non-Gaussian signal from single-field, slow-roll inflation is
suppressed by slow-roll parameters to an undetectable level
\cite{Maldacena:2002vr}, if non-Gaussianities are detected then this
would rule out many models of inflation.

Thus, as a natural extension of the study performed in \cite{KW,KMW},
in this paper
we compute the non-Gaussianity of the primordial
curvature perturbations generated from the contracting
phase of the multi-field new ekpyrotic cosmology.
We adopt the same specific model as the previous analysis of
Refs.~\cite{KW,KMW} where the scale-invariant curvature perturbation
is generated by the transition from a scaling solution, with two
fields driving the collapse, to a single-field dominated regime.
We assume that the ekpyrotic collapse is subsequently converted to
expansion by a regular bounce, during which the comoving curvature
perturbation is conserved on large scales and the curvature
perturbation generated during the collapse is thus directly related
to the amplitude of the observed primordial density perturbation.

This paper is organized as follows. In Sec.~\ref{sec2} we briefly
review the background dynamics. In Sec.~\ref{sec3} we summarize the
statistical properties of linear and non-linear perturbations. The
$\delta N$-formalism is introduced which is used to compute the
primordial curvature perturbation. In Sec.~\ref{sec4} we study 
linear perturbations while in
Sec.~\ref{sec5} we generalize this study to non-linear perturbations
and quantify the expected non-Gaussianity. In Sec.~\ref{sec6} we
draw our conclusions. In three appendices we review previous results
for the linear fluctuations of the
isocurvature field during the ekpyrotic phase and calculate its
intrinsic non-Gaussianity using the interaction Hamiltonian, as well as
presenting a numerical check of our analytical results.

\section{Homogeneous dynamics}
\label{sec2}

We first review the background dynamics of the fields in the new
ekpyrotic cosmology with multiple scalar fields. During the
ekpyrotic collapse the contraction of the universe is assumed to be
described by a 4D Friedmann equation in the Einstein frame with $n$
scalar fields with negative exponential potentials
 \begin{equation}
 3 H^2 = V + \sum_j ^n \frac12 \dot\phi_j^2 \,,
 \end{equation}
where
\be
 \label{Vi}
 V = - \sum_j ^n V_j e^{-c_j\phi_j} \,,
\ee
and we take $V_i>0$
and set $8\pi G$ equal to unity.

From now on, for simplicity, we concentrate our attention on the
case of two fields. In this case, it will be easier to work in terms
of new variables \cite{KW},
\be
 \varphi = \frac{c_2 \phi_1 + c_1 \phi_2}{\sqrt{c_1^2+c_2^2}} \,,
\quad \chi = \frac{c_1 \phi_1 - c_2 \phi_2}{\sqrt{c_1^2+c_2^2}} \,,
\label{sigmachi}
 \ee
corresponding to a fixed rotation in field space.
The potential given by
Eq.~(\ref{Vi}) can then be simply re-written as
\cite{Finelli:2002we,KW,Malik:1998gy}
 \be
 \label{Vsigmachi}
 V = - U(\chi) \, e^{-c\varphi} \,,
 \ee
where
 \be
  \frac{1}{c^2} \equiv \sum_j \frac{1}{c_j^2}\,, \label{c}
 \ee
and the potential for the orthogonal field is given by
 \be U(\chi) =
V_1\, e^{-(c_1/c_2)c\chi} + V_2\, e^{(c_2/c_1)c\chi} \,,
\label{U_chi}
 \ee
which has a minimum at
 \be
  \chi=\chi_0 \equiv
\frac{1}{\sqrt{c_1^2+c_2^2}} \ln \left(\frac{c_1^2 V_1}{c_2^2
V_2}\right)\,.
 \ee
If we expand $U(\chi)$ in Eq.~(\ref{U_chi}) about its minimum we
obtain
 \be
  U(\chi) = U_0 \left[1 + \frac{c^2}{2} (\chi - \chi_0)^2 +
\frac{\tilde
  c c^2}{6} (\chi - \chi_0)^3 + \ldots \right],
 \ee
where
 \be
  \tilde c \equiv \frac{c_2^2 - c_1^2}{\sqrt{c_1^2 +
c_2^2}}. \label{tildec}
 \ee

The multi-field scaling solution corresponds
to the classical solution
along this minimum $\chi =\chi_0$,
while $\varphi$ is rolling down the exponential potential.
The explicit form of the multi-field scaling solution
is given as
 \begin{eqnarray}
&&a=(-t)^p\,,\label{ev_a_multi_scaling}\\
&&\varphi= \frac{2}{c} \ln (-t)-\frac{1}{c}
\ln \left( \frac{p(1-3p)}{U_0}\right)\,,
\label{ev_vphi_multi_scaling}
 \end{eqnarray}
where $p= \sum_j 2/c_j^2=2/c^2$. The potential for $\chi$ has a
negative mass-squared around $\chi =\chi_0$,
 \be m_\chi^2 \equiv
\frac{\partial^2 V }{\partial \chi^2} = c^2 V <0 \,,
\label{m_chi}
 \ee
and thus $\chi$ represents the instability direction. Furthermore, 
the $\chi$ field evolution is nonlinear, with
the cubic interaction being given by
 \begin{equation}
V^{(3)} \equiv \frac{\partial^3 V }{\partial \chi^3} =  \tilde c
m_\chi^2\,, \label{V3}
 \end{equation}
which becomes important when we consider the non-Gaussianity later
in this paper.

If the initial condition for $\chi$ is slightly different from $\chi_0$ or
$\dot{\chi}$ is not zero, then $\chi$ starts rolling down the potential
and the solution approaches a single-field-dominated scaling
solution.
The explicit form of the single-field-dominated
scaling solution is given as
 \begin{eqnarray}
&&a=(-t)^{p_j}\,,\label{ev_a_sing_scaling}\\
&&\phi_j= \frac{2}{c_j} \ln (-t)-\frac{1}{c_j}
\ln \left( \frac{p_j(1-3p_j)}{V_j}\right)\,,
\label{ev_phi_sing_scaling}
 \end{eqnarray}
where $p_j= 2/c_j^2$. In this paper, we consider the case in which
the background evolves from the multi-field scaling solution to the
$\phi_2$-dominated scaling solution without loss of generality.

\section{Statistical correlators}
\label{sec3}

Here we briefly summarize the statistical properties of the scalar
field fluctuations during ekpyrotic collapse. Then, in order to link
these to the observable primordial curvature perturbation we discuss
the $\delta N$-formalism.

In the two-field new ekpyrotic cosmology, the isocurvature
fluctuations acquired by the field $\chi$ during the multi-field
scaling regime, play a crucial role to generate a scale-invariant
spectrum of perturbations. On the other hand, the fluctuations of
the field $\varphi$ are negligible on large scales, because of its
very blue spectral tilt
\cite{Lehners:2007ac,Buchbinder:2007ad,Creminelli:2007aq}. Thus, in
the following we neglect $\delta \varphi$ fluctuations.

In linear perturbation theory, when interactions are neglected,
the free-field fluctuations $\delta \chi$ are Gaussian. Their
power spectrum ${\cal P}_\chi$ can be defined by
\begin{equation}
\langle \delta \chi_{{\bf k}_1} \delta \chi_{{\bf k}_2} \rangle
\equiv (2 \pi)^3 \delta^3({{\bf k}_1} + {{\bf k}_2})\frac{2
\pi^2}{k_1^3} {\cal{P}}_{\chi}(k_1)\,, \label{power_spectrum_scalar}
\end{equation}
where the angle brackets denote an ensemble average.

If the cubic self-interaction in Eq.~(8) is taken into account 
$\delta \chi$
is no longer Gaussian and 
the first signal of non-Gaussianity comes from the three-point
correlation function. Similarly to
Eq.~(\ref{power_spectrum_scalar}), the bispectrum of $\delta \chi$,
$B_\chi$, is defined by
\begin{equation}
\langle \delta \chi_{{ \bf k}_1} \delta \chi_{{ \bf k}_2} \delta
\chi_{{ \bf k}_3} \rangle \equiv  (2 \pi)^3 \delta^3(\sum_j {\bf
k}_j )B_{\chi} (k_1, k_2, k_3) \,. \label{scalar_bispectrum_def}
\end{equation}

To characterize the bispectrum one can also 
define the nonlinear parameter
of the field fluctuation, $f_{NL}^\chi$, as
\begin{equation}
\frac{6}{5} f_{NL}^\chi \equiv
\frac{\prod_j k_j^3}{\sum_j k_j^3}
\frac{B_{\chi}}{4 \pi^4 {\cal{P}}_\chi^2}\,.
\label{fnl_chi_def}
\end{equation}
When the intrinsic non-Gaussianity is local in real
space this parameter is $k$ independent and $\delta \chi$ can be written as
\begin{equation}
 \delta\chi = \delta\chi_L + \frac35 f_{NL}^\chi \delta \chi_L^2 \,,
\label{fnl_chi}
\end{equation}
where $\delta \chi_L$ is the linear and Gaussian part of the field fluctuations.

To relate the non-Gaussianity of the scalar field fluctuations to
observations, we need to calculate the three-point functions of the
comoving curvature perturbation $\zeta$. In order to do that, we can
use the $\delta N$-formalism
\cite{Starobinsky:1986fx,Sasaki:1995aw,Sasaki:1998ug,Vernizzi:2006ve,Lyth:2005fi,Byrnes:2007tm}. In the $\delta N$-formalism, the comoving curvature
perturbation $\zeta$ evaluated at some time $t=t_f$ coincides with
the perturbed expansion integrated from an initial $flat$
hypersurface at $t=t_i$, to a final {\it uniform density}
hypersurface at $t=t_f$, with respect to the background expansion,
i.e.,
\begin{equation}
\zeta(t_f, {\bf x}) \simeq \delta N(t_f, t_i, {\bf x}) \equiv
\N(t_f, t_i, {\bf x})- \barf{N}(t_f, t_i)\,,
\label{rel_zeta_deltan}
\end{equation}
with
\begin{equation}
\N(t_f, t_i, {\bf x}) \equiv \int_{t_i} ^{t_f} {\cal H}({\bf x},t)
dt\,, \;\;\;\; \barf{N}(t_f, t_i) \equiv \int_{t_i} ^{t_f}
\barf{H}(t) dt\,, \label{expansion_def}
\end{equation}
where ${\cal H}({\bf x},t)$ is the inhomogeneous Hubble expansion.
We can calculate $\delta N$ on large scales using the homogeneous
equations of motion, in the assumption that the local expansion on
sufficiently large scales behaves like a locally homogeneous and
isotropic universe, according to the so-called ``separate universe''
approach \cite{Salopek:1990jq,Wands:2000dp,Lyth:2003im}. This allows us to compute
the full nonlinear curvature perturbation in the large-scale limit.
Note that we leave the initial time $t_i$ unspecified and we are
free to identifying it with the time of Hubble crossing $t=t_*$,
i.e. the time when a mode $k$ exits the Hubble radius during
inflation, $k=aH$, or with a later time.

We will chose the initial time $t_i$ to be {\em during} the
multi-field scaling regime. Furthermore, since $\varphi$ is
unperturbed, $\delta N$ can be expanded in series of the initial
field fluctuations $\delta \chi_i$. Retaining only terms up to
second order, we obtain
\begin{equation}
\delta N  = \barf{N}_{,\chi_i}\delta \chi_i + \frac{1}{2}
\barf{N}_{,\chi_i \chi_i} (\delta \chi_i)^2\,, \label{deltan_exp}
\end{equation}
where $\barf{N}_{,\chi}$ denotes the derivative of $\barf{N}$ with
respect to $\chi$.

Now we can convert the higher-order information
about the initial field fluctuations
into the statistical properties of
the observed primordial curvature perturbations.
The power spectrum of the comoving curvature perturbation $\zeta$,
${\cal{P}}_\zeta$, is defined as
\begin{equation}
\langle \zeta_{{\bf k}_1} \zeta_{{\bf k}_2} \rangle
\equiv (2 \pi)^3 \delta^{(3)} ({{\bf k}_1}+{{\bf k}_2})
\frac{2 \pi^2}{k_1^3} {\cal{P}}_\zeta(k_1)\,.
\label{power_spectrum_curvature}
\end{equation}
At lowest order, from Eqs.~(\ref{power_spectrum_scalar}),
(\ref{deltan_exp}) and (\ref{power_spectrum_curvature}),
${\cal{P}}_\zeta$ is expressed as
\begin{equation}
 {\cal{P}}_\zeta  = \barf{N}_{,\chi_i}^2
{\cal{P}}_{\chi_i} \,.
\label{rel_zeta_chi_powerspectrum}
\end{equation}
Note that $\barf{N}_{,\chi_i}$ is independent of wavenumber $k$ and
hence the scale dependence of the primordial spectrum, $\Delta
n\equiv d\ln{\cal{P}}_\zeta/d\ln k$, is given by the spectral tilt
of the field fluctuations, $\Delta n_\chi$, on the initial
hypersurface [given in Eq.~(\ref{tilt}) in
Appendix~\ref{appendix1}].

The bispectrum of the curvature perturbation $\zeta$, which includes
the first signal of non-Gaussianity, is defined as
\begin{equation}
\langle \zeta_{{\bf k}_1} \zeta_{{\bf k}_2} \zeta_{{\bf k}_3} \rangle
\equiv (2 \pi)^3 \delta^{(3)} (\sum_j {{\bf k}_j})
B_\zeta (k_1, k_2, k_3)\,,
\label{bispectrum_curv_def}
\end{equation}
where the left hand side of Eq.~(\ref{bispectrum_curv_def})
can be evaluated by the $\delta N$-formalism using Wick's theorem,
\begin{equation}
\langle \zeta_{{\bf k}_1} \zeta_{{\bf k}_2} \zeta_{{\bf k}_3}
\rangle = \barf{N}_{,\chi_i}^3 \langle \delta \chi_{i{\bf k}_1}
 \delta \chi_{i{\bf k}_2}  \delta \chi_{i{\bf k}_3} \rangle
+\frac12  \barf{N}_{,\chi_i}^2 \barf{N}_{,\chi_i \chi_i} \langle
\delta \chi_{i{\bf k}_1} \delta \chi_{i{\bf k}_2} (\delta \chi_i
\star \delta \chi_i)_{{\bf k}_3} \rangle +{\rm perms}\,. \label{perms}
\end{equation}
In the above equation, a star $\star$ denotes the convolution and
we have neglected correlators higher than the
four-point.

Observational limits on the non-Gaussianity of the primordial
curvature perturbations are usually given on the nonlinear parameter
$f_{NL}$ defined by \cite{Maldacena:2002vr}
\begin{equation}
\frac{6}{5} f_{NL} \equiv
\frac{\prod_j k_j^3}{\sum_j k_j^3}
\frac{B_{\zeta}}{4 \pi^4 {\cal{P}}_\zeta^2}\,.
\label{fnl_def}
\end{equation}
If the non-Gaussianity is local, one can write
$\zeta$ as
\begin{equation}
 \delta N = \zeta_L + \frac35 f_{NL} \zeta_L^2 \,, \label{fnl_local}
\end{equation}
where $\zeta_L$ is a Gaussian variable.

To compute the bispectrum of the curvature perturbation one can use
the $\delta N$-formalism and after some manipulations,
from Eqs.~(\ref{bispectrum_curv_def})
and (\ref{perms}) one finds
\begin{equation}
B_\zeta (k_1,k_2,k_3) = \barf{N}_{,\chi_i}^3 B_{\chi_i} (k_1,k_2,k_3)+ 4
\pi^4 {\cal{P}}_\zeta^2 \frac{\sum_j k_j^3}{\prod_j k_j^3}
\frac{\barf{N}_{,\chi_i \chi_i}} {\barf{N}_{,\chi_i}^2}\,,
\end{equation}
where $B_{\chi_i} (k_1,k_2,k_3)$ is the bispectrum of
the scalar field evaluated at $t=t_i$.
Together with Eq.~(\ref{fnl_def}) the nonlinear
parameter $f_{NL}$ becomes
\begin{equation}
f_{NL} =
\frac{f_{NL}^{\chi_i} }{N_{,\chi_i}}
+
\frac{5}{6}  \frac{\barf{N}_{,\chi_i \chi_i}}
{\barf{N}_{,\chi_i}^2} \,, \label{fnl_curvature}
\end{equation}
where we have used Eqs.~(\ref{fnl_chi}) and (\ref{fnl_def}).
The first term on the right hand side
of Eq.~(\ref{fnl_curvature}),
\begin{equation}
f_{NL}^{(3)} \equiv \frac{f_{NL}^{\chi_i} }{N_{,\chi_i}}
\,,
\label{fnl3_def}
\end{equation}
comes from the intrinsic three-point correlation functions
of the field fluctuations
and contains also the non-Gaussianity of {\em quantum} origin
generated on small-scales, i.e.,
{\it inside}
the Hubble radius during the ekpyrotic collapse
described by the multi-field scaling solutions.
The second term
on the right hand side
of Eq.~(\ref{fnl_curvature}),
\begin{equation}
f_{NL}^{(4)} \equiv \frac{5}{6}
 \frac{\barf{N}_{,\chi_i \chi_i}}{\barf{N}_{,\chi_i}^2}\,,
\label{fnl4_def}
\end{equation}
is completely momentum independent and
{\it local} in real space, because it is due to
the evolution of nonlinearities {\it outside}
the Hubble radius during the multi-field
ekpyrotic collapse. Note that the splitting in $f_{NL}^{(3)}$ and $f_{NL}^{(4)}$
depends on the time $t_i$.

\section{Linear curvature perturbation}
\label{sec4}

The power spectrum of $\chi$ is given by (see \cite{KMW} and
Eq.~(\ref{power_delchi}) in Appendix A)
\begin{equation}
{\cal{P}}_\chi (k)  = \epsilon^2 \left( \frac{H}{2 \pi}\right)^2\,,
\label{PSchi}
\end{equation}
in the limit of large $\epsilon$,
where the fast-roll parameter $\epsilon$ is defined as
$\epsilon
\equiv - \dot{H}/{H^2} = c^2/2=1/p$, and it is constant for the
multi-field scaling solution. Thus, in this limit, ${\cal{P}}_\chi$
has a scale invariant
spectrum.

On the other hand, as shown in \cite{KMW}
in the fast-roll limit and assuming an {\em instantaneous
transition} from the multi-field scaling solution to the
single-field $\phi_2$-
dominated scaling solution, the power spectrum of the
final (after the transition) curvature perturbations becomes
\begin{equation}
{\cal P}_\zeta =  \frac{\epsilon^2}{c_1^2+c_2^2} \left|
\frac{H_T}{2 \pi}  \right|^2 \,, \label{power_zeta_ht_2}
\label{powerspec}
\end{equation}
where $H_T$ is the Hubble parameter evaluated
at the transition time $t=t_T$.
The scalar spectral index of $\zeta$ is 
\begin{equation}
\Delta n = 4 \left(\frac{1}{c_1^2}
+\frac{1}{c_2^2}\right)\,.
\label{scal_spec_index}
\end{equation}
Note that it is always positive, i.e., the power spectrum
is always blue

Let us now interpret the result above in the light of the $\delta N$-formalism.
We consider the situation in which $\chi_i$ is perturbed on the
$t=t_i$ hypersurface, while $H_i$ assumes on this hypersurface a
constant value. This is justified by the fact that the $t=t_i$
hypersurface is flat and since $\chi$ is an isocurvature field its
fluctuations do not affect the local Hubble expansion.
Furthermore, we assume that the transition into the single-field-dominated
scaling solution at the time $t=t_T$, happens {\em instantaneously}
on the hypersurface
$\chi=\chi_T=$ const., where $H_T$ is perturbed.

Under these assumptions,
the expansion $\barf{N}$ defined by
Eq.~(\ref{expansion_def}) can be split into
\begin{equation}
\barf{N} = \int_{t_i} ^{t_T} \barf{H} dt + \int_{t_T} ^{t_f}
\barf{H} dt\,, \label{expansion_splitted_2}
\end{equation}
where $t_f$ is set sufficiently later than the transition time
$t_T$. In Eq.~(\ref{expansion_splitted_2}), 
the first integral is over 
the multi-field scaling evolution and the last integral is over the 
$\phi_2$-dominated phase.

Since the multi-field scaling solution is characterized by
Eq.~(\ref{ev_a_multi_scaling}), the first
term on the right hand side of
Eq.~(\ref{expansion_splitted_2}) can be expressed as
$(1/\epsilon) \ln (\barf{H}_i/\barf{H}_T)$.
Similarly, since the single-field dominated scaling solution
is characterized by Eq.~(\ref{ev_a_sing_scaling}),
the second term becomes
$(1/\epsilon_2) \ln (\barf{H}_T/\barf{H}_f)$,
where $\epsilon_2=c_2^2/2$. Then, for
a fixed $t_i$ and $t_f$, the expansion $\barf{N}$ can be expressed
as
\begin{equation}
\barf{N} = -\frac{2}{c_1^2} \ln |\barf{H}_T| + {\rm const.},
\label{expansion_ito_ht_2}
\end{equation}
which depends only on the parameter $c_1$,
besides the transition time $t_T$.

During the multi-field scaling regime, the linear evolution equation
of $\chi$ on large scales is given by \be \ddot \chi + 3 H \dot \chi
+ m_\chi^2 \chi =0\,,  \label{homo_eq} \ee where the mass of $\chi$
is defined in Eq.~(\ref{m_chi}). During the multi-field scaling the
evolution of $\chi$ is dominated by the tachyonic mass and thus $3 H
\dot \chi \ll m_\chi^2 \chi$. With $m_\chi^2 = -2/t^2$ (see the
appendix, Sec.~\ref{appendix1}), one finds $\chi \propto 1/t \propto
H$, and thus \be H_T = H_i \frac{\chi_T}{\chi_i}. \label{linear_rel}
\ee Using this relation one can derive $N$ with respect to $\chi_i$
and obtain \be N_{,\chi_i} =  \frac{2}{c_1^2 \chi_i} = 
\frac{2}{c_1^2 \chi_T} \frac{H_T}{H_i}. \label{rel_dndchi_hthi} \ee
In particular, by using this equation and comparing
Eq.~(\ref{rel_zeta_chi_powerspectrum}) with
Eq.~(\ref{power_zeta_ht_2}), one obtains the value of $\chi_T$,
i.e.,
\be \chi_T = \frac{2 \sqrt{c_1^2 + c_2^2}}{c_1^2}. \label{chi_T} \ee
In appendix C, we have calculated $N_{,\chi_i}$ numerically and
checked Eq.~(\ref{chi_T}).

\section{Non-Gaussianities}
\label{sec5}

In this section we compute the non-Gaussianity of the primordial
curvature perturbation generated by the nonlinear dynamics during
multi-field ekpyrotic collapse. We will use the $\delta N$-formalism
to calculate $\delta N(\chi)$ in the instantaneous transition
approximation. We will thus consider quadratic terms in $\delta
\chi_i$ in the $\delta N$-expansion (\ref{deltan_exp}) and, in
contrast with Sec.~\ref{sec4}, we also allow for non-linear
evolution of $\chi$ on large scales.
For another example of the multi-field calculation
of non-Gaussianity from fields with exponential potential
see \cite{Seery:2005gb}, 
even though in the context of assisted inflation.

Including the cubic self-interaction $V^{(3)}$ given in
Eq.~(\ref{V3}), the large scale evolution equation for $\chi$ in the
multi-field scaling regime becomes \be \ddot \chi + 3 H \dot \chi +
m_\chi^2 \chi = -\frac{1}{2} \tilde c m_\chi^2 \chi^2. \ee The above
evolution equation can be solved perturbatively. Given the solution
to the linear equation (\ref{homo_eq}), i.e., $\chi_L \propto H$,
the growing-mode solution for $\chi$ is
\begin{equation}
 \chi = \chi_L + \frac14 \tilde{c} \chi_L^2=\alpha H + \frac14 \tilde{c}\alpha^2 H^2 \,, \label{chi_nl}
\end{equation}
where $\alpha$ is a
constant parameter whose value distinguishes the different
trajectories.
Note that this is a {\em perturbative} result, i.e., it is valid
only as long as $\tilde c \chi \ll 1 $. However, since $\chi$ grows
during the collapse,  unless
prevented by a bouncing phase, eventually this condition is violated.

For perturbations in the value of $\chi$ on a hypersurface of uniform $H$
we have
\begin{equation}
 \label{NLdeltachi}
 \delta\chi = \left( 1 + \frac12 \tilde{c}\alpha H \right) H \delta\alpha + \frac14 \tilde{c}H^2 (\delta\alpha)^2 \,.
\end{equation}
The non-linear self-interaction of $\chi$ in Eq.~(41) 
grows in time on super-Hubble scales, 
and thus the intrinsic non-Gaussianity of $\chi$ increases. 
Therefore, unless we take $t_{\rm i}$ sufficiently early
that we can neglect these nonlinearities, 
we cannot naively use Eq.~(39), 
which is only valid at linear order, 
and its derivative with respect to $\chi_i$, 
to estimate the non-Gaussianity of the curvature perturbation. 
We need to work with a variable that is as close as 
possible to a Gaussian. 
It turns out that it is convenient to choose $\alpha$ 
as such a variable.
Indeed, if we assume that $\delta \chi_L$, 
the perturbation of the solution
of the linear equation (\ref{homo_eq}), is Gaussian, then also
$\delta\alpha$ is a Gaussian random variable because $ \delta \alpha
= \delta \chi_L/H $. Comparing this equation with
Eq.~(\ref{fnl_chi}) we find that the intrinsic non-Gaussianity of
$\chi$ is of local form and time independent, and it is given by
 \be
f_{NL}^\chi=\frac{5}{12}\tilde{c}. \label{fnl_chi_m}
 \ee
In Sec.~\ref{app2} of the appendix, we have checked that this result
agrees with the one obtained using the approach of Maldacena
\cite{Maldacena:2002vr} with an interaction Hamiltonian containing
$V^{(3)}$ \cite{Creminelli:2007aq}, and confirms that we can
consistently assume $\delta \chi_L$, and thus $\delta \alpha$, to be
Gaussian. The reason for this is that the classical nonlinear
evolution on large scales completely dominates over the sub-Hubble
nonlinear interactions of quantum nature.

We will now compute the non-Gaussianity of the curvature
perturbation using Eqs.~(\ref{fnl_curvature}--\ref{fnl4_def}). The
final result for $f_{NL}$ will be manifestly independent of the
time we choose to evaluate $\delta\chi_i$ although the splitting
into $f_{NL}^{(3)}$ and $f_{NL}^{(4)}$ is dependent on $t_i$.

Since $\delta \alpha$ can be assumed to be Gaussian, the simplest
way to compute $f_{NL}$ is to calculate the $\delta N$
corresponding to the fluctuation $\delta\alpha$, i.e.,
\begin{equation}
 \delta N = N_{,\alpha} \delta\alpha + \frac12 N_{,\alpha\alpha} (\delta\alpha)^2
 \,.
\end{equation}

In order to compute $N_{,\alpha}$ and $N_{,\alpha\alpha}$ we want
to use Eq.~(\ref{expansion_ito_ht_2}), and for this we need to
know how $H_T$ varies as a function of $\alpha$ at the transition
from multi-field scaling to single-field
$\phi_2$-dominated scaling solution.
Inverting Eq.~(\ref{chi_nl}) (to leading order in $\tilde{c}
\chi$) gives
\begin{equation}
 \label{alphachi}
 \alpha = \frac{\chi}{H} \left( 1 - \frac14 \tilde{c}\chi \right) \,.
\end{equation}
Assuming as in the linear case that the transition corresponds to
a critical value of the tachyon field $\chi=\chi_T$, on the
transition surface (constant $\chi_T$) we have from
(\ref{alphachi}) that $\alpha\propto H_T^{-1}$ and hence we find
\begin{equation}
 \label{deltaNdeltaalpha}
 \delta N = \frac{2}{c_1^2} \frac{\delta\alpha}{\alpha}
  - \frac{1}{c_1^2} \left( \frac{\delta\alpha}{\alpha} \right)^2 \,,
\end{equation}
which means
\begin{equation}
 N_{,\alpha} =  \frac{2}{c_1^2} \frac1 \alpha,\qquad
 N_{,\alpha \alpha} = - \frac{2}{c_1^2} \frac{1}{\alpha^2} \,.
\end{equation}

Taking $\delta\alpha$ to be a Gaussian random variable
and comparing with Eq.~(\ref{fnl_local}) with $\zeta_L=-2 \delta
\alpha /(c_1^2 \alpha)$
we obtain the nonlinear parameter for the curvature perturbation after
the transition:
\begin{equation}
 \label{totalfNL}
 f_{NL} = \frac{5}{6} \frac{ N_{,\alpha \alpha}}
{ N_{,\alpha}^2}= -\frac{5}{12} c_1^2 \,.
\end{equation}
This is our main result. The non-Gaussianity is given in terms of
$c_1$, where $-V_1 \exp (-c_1 \phi_1)$ is the potential of the field $\phi_1$
which remains subdominant after the transition from the multi-field
scaling to the single-field dominated regime. The
non-Gaussianity is of local form. This is due to the fact that 
it is generated by the nonlinear super-Hubble evolution.

Equation (\ref{totalfNL}) includes also the non-linear growth of
the tachyon field on large scales due to its self-interaction. In
order to see this, we can compute $f_{NL}^{(3)}$ and
$f_{NL}^{(4)}$ defined in Eqs.~(\ref{fnl3_def}) and
(\ref{fnl4_def}). Thus, we have to identify the linear and
non-linear dependence of $\delta N$ on the field values
$\delta\chi_i$ on the initial hypersurface.

Replacing Eq.~(\ref{fnl_chi_m}) in Eq.~(\ref{fnl3_def})
we obtain
\begin{equation}
 f_{NL}^{(3)} = \frac{5}{12} \frac{\tilde{c}}{N_{,\chi_i}}   \,.
\end{equation}
Furthermore, from Eqs.~(\ref{alphachi})  and (\ref{deltaNdeltaalpha})
we have
\begin{equation}
 N_{,\chi_i} = \frac{dN}{d\alpha} \frac{d\alpha}{d\chi_i} =  \frac{2}{c_1^2\chi_i} \left( 1 - \frac14 \tilde{c}\chi_i \right) \,,
\end{equation}
and hence
\bea
 f_{NL}^{(3)}   &=& \frac{5}{24} c_1^2 \tilde{c} \chi_i  \\ &=&  \frac{5}{12} (c_2^2 - c_1^2)\frac{H_i}{H_T}  \,,
\label{res_fnl3}
\eea
where for the last equality we have replaced $\tilde c$ using its
definition, Eq.~(\ref{tildec}), expressed $\chi_i$ in terms of
$\chi_T$ using the linear relation Eq.~(\ref{linear_rel}), and
replace $\chi_T$ using Eq.~(\ref{chi_T}).

Secondly we have the contribution due to
\begin{equation}
 N_{,\chi_i \chi_i } = \frac{d^2N}{d\alpha^2} \left( \frac{d\alpha}{d\chi_i} \right)^2 + \frac{dN}{d\alpha} \frac{d^2\alpha}{d\chi_i^2} =  -\frac{2}{c_1^2\chi_i^2} \,.
\end{equation}
Note that this relation is valid to linear order in $\tilde c
\chi_i$, and thus contains also the nonlinear self-interaction of
$\chi$ generating non-Gaussianities after $t_i$. Using
Eq.~(\ref{fnl4_def}),
this gives
\bea
 f_{NL}^{(4)} &=&  -\frac{5}{12} c_1^2 \left( 1 + \frac12
\tilde{c}\chi_i \right) \\ &=& -\frac{5}{12} c_1^2 - \frac{5}{12}
(c_2^2 - c_1^2)\frac{H_i}{H_T}\,. \label{f4}
\label{res_fnl4}
\eea
Both $f_{NL}^{(3)}$ and $f_{NL}^{(4)}$ depend upon the choice of
the initial hypersurface and thus of $\chi_i$. However the total
$f_{NL}$ comes from the sum of the two terms, it is independent of
$t_i$, and is given by Eq.~(\ref{totalfNL}).

In appendix C, we calculated  $N_{,\alpha}$ and $N_{,\alpha \alpha}$
numerically and confirm that the result Eq.~(\ref{totalfNL})
obtained by the instantaneous transition
and the fast roll approximations is satisfied with
good accuracy.



\section{Conclusion}
\label{sec6}

In this paper we have studied the nonlinear evolution of
perturbations in the multi-field new ekpyrotic cosmology. If one
sets the model parameters and initial conditions appropriately, then
the prediction for the power spectrum of curvature perturbations
produced in multi-field ekpyrotic collapse can be 
constrained by present observations
\cite{Lehners:2007ac,Buchbinder:2007ad,Creminelli:2007aq,KMW}. In
order to distinguish this model from other early universe scenarios,
such as inflation, by future observations, it is important to
estimate the non-Gaussianity of the curvature perturbations.

We have studied the simplest model based on two fields with
exponential potentials and considered the specific scenario in which
the nearly scale-invariant comoving curvature perturbation is
generated by the transition from the multi-field scaling solution to
the single-field dominated attractor solution.
We have applied the $\delta N$-formalism, which is widely adopted
to study the non-linearity of the primordial curvature perturbation
from inflation \cite{Lyth:2005fi}, to ekpyrotic cosmology. We
identify the non-linear curvature perturbation on uniform-density
hypersurfaces at late times with the perturbed local expansion,
$\delta N$, with respect to an initial spatially flat hypersurface.
The primordial non-Gaussianity parameter $f_{NL}$ is a sum of two
contributions: $f_{NL}^{(3)}$, defined in Eq.~(\ref{fnl3_def}),
comes from the intrinsic three-point function of the isocurvature
field perturbation $\delta \chi$ on an initial hypersurface, and
$f_{NL}^{(4)}$, defined in Eq.~(\ref{fnl4_def}), originates from the
nonlinear relation between the primordial curvature perturbation and
the isocurvature field perturbations during multi-field ekpyrotic
collapse.
It should be emphasized that although the decomposition of $f_{NL}$
into $f_{NL}^{(3)}$ and $f_{NL}^{(4)}$ 
may be convenient, it is unphysical and depends upon the choice of the initial
time $t_i$. However, we show that the physical quantity, the total $f_{NL}$,
is independent of $t_i$.

Both the multi-field and single-field ekpyrotic solutions are
power-law solutions. We find a general result that in case of a
sudden transition between two power-law solutions the local
expansion is only a function of the Hubble rate at the transition,
$H_T$. Thus the calculation of the primordial curvature perturbation
in our model reduces to finding the perturbation of $H_T$ for
different trajectories in phase-space, and hence the local value of
the isocurvature field $\chi$ on the initial spatially flat
hypersurface.


We find that after the transition to the single-field attractor
solution the non-Gaussian parameter $f_{NL}=-5c_1^2/12$, where $-V_1
\exp (-c_1 \phi_1)$ is the potential of the field $\phi_1$ which
becomes subdominant at late time. Since the non-Gaussianity is
mainly generated by the 
nonlinear
super-Hubble evolution, it is of the local form, and the nonlinear parameter
is $k$ independent.
We show in appendix \ref{app2} that the contribution of the
intrinsic non-Gaussianity of the isocurvature field perturbations on
sub-Hubble scales is subdominant.

We have checked our analytical results by calculating the expansion,
$N$, numerically in appendix \ref{app3}. We confirmed that when the
fast-roll parameter satisfies $ 400 > \epsilon > 25$ the analytic
estimation is accurate within at a few $\%$ level. The discrepancy
arises from the breakdown of both the fast-roll approximation and
the sudden transition approximation.

A negative value of $f_{\rm NL}$ is much more tightly constrained by
current observations than a positive value. For instance, if we
choose $c_1 = 5$ we obtain a nonlinear parameter
$f_{\rm NL}\approx-10$
that is marginally consistent with current constraints on the
non-Gaussianity of the primordial density perturbation
\cite{Komatsu:2008hk,Smith:2009jr}. However such a small value of $ c_1$ leads to
too large a value of the spectral index (35), $\Delta n > +0.16$,
which is excluded by observations \cite{Komatsu:2008hk}.

Corrections to the exact exponential potentials that we have studied
in this paper have been proposed
\cite{Lehners:2007ac,Buchbinder:2007ad} to produce a red spectrum of
perturbations. Indeed corrections are also required to stabilise the
multi-field scaling solution at early times (before observables
scales exit the Hubble scale) \cite{Buchbinder:2007tw}. We assume
that the classical background solution starts close to the
multi-field scaling solution. However, as the multi-field scaling
solution is a saddle point in the phase space, we need some
preceding phase that initially drives the classical background
solution to the saddle point. It is important to check whether such
corrections will also affect the non-linear evolution of the field
perturbations during the collapse phase, which could lead to
additional sources of primordial non-Gaussianity.

Modifications to the effective action are certainly required at high
energies to turn contraction to expansion before the collapse phase
reaches the big crunch singularity
\cite{Creminelli:2006xe,Lehners:2007ac,Buchbinder:2007ad,Creminelli:2007aq}.
In this paper, we assumed that the comoving curvature perturbation
is conserved on super-Hubble scales through such a non-singular
bounce, neglecting the effect of non-adiabatic perturbations which
are expected to rapidly decay about the single-field attractor
solution. In this case the power spectrum and the non-Gaussian
parameter Eq.~(\ref{totalfNL}) that we have calculated in the
single-field dominated collapse are directly related to those
observed perturbations in the expanding, hot big bang phase.
It is certainly possible to convert the initial isocurvature field
perturbations to curvature perturbations through a different
mechanism, e.g., via the bounce
\cite{Lehners:2007ac,Buchbinder:2007ad,Creminelli:2007aq,Buchbinder:2007tw}.
In this case the non-linear dependence of the expansion, $\delta N$,
upon the initial field perturbations may be different, but the
non-linear growth on super-Hubble scales of the isocurvature field
perturbations about the multi-field scaling solution is still
expected to lead to a large non-Gaussianity
\cite{Creminelli:2007aq,Buchbinder:2007tw}.
Further work is required to quantify the non-Gaussianity predicted in
these models.

{\em Note added: This arXiv version of our article includes corrections
made after publication of the journal article. 
We give the correct expressions for several equations that have an
incorrect sign. This incorrect sign is due to a sign error in the
integrated expansion $N$ 
in Eq.~(37), which then propagates into
other equations. As a consequence, the nonlinear parameter $f_{\rm
NL}$ in this new ekpyrotic scenario turns out to be negative and it
is much more tightly constrained by current data than it would have
been with a positive sign. By combining the present observational
constraints on $f_{\rm NL}$ and the scalar spectral index, the
specific model studied in this paper is thus ruled out by current
observational data.}

\section*{Acknowledgments}
We are grateful to P. Creminelli and M. Sasaki for helpful
discussions. KK and DW are supported by STFC. SM is supported by a
JSPS Research Fellowship. SM is grateful to the ICG, Portsmouth, for
their hospitality when this work was initiated. This work is
supported by JSPS, Japan-U.K.~Research Cooperative Program.
We thank Jean-Luc Lehners and Paul Steinhardt 
for useful discussions which 
lead us to discover this sign error. 
KK is grateful to the Princeton Center for 
Theoretical Science where part of this work was done.

\appendix
\section{Linear field fluctuations}
\label{appendix1}

Here, we briefly summarise the results obtained in the previous
works
\cite{Lehners:2007ac,Buchbinder:2007ad,Creminelli:2007aq,Finelli:2002we,KW}
about the generation and the evolution of linear fluctuations of the
$\chi$ field about the background multi-field scaling solution.

In the multi-field scaling solution $\delta \chi$ 
coincides with entropy field perturbation, which is
automatically gauge-invariant and does not couple with
gravity at first order.
Since there is no coupling between
the adiabatic and the entropy field
perturbations, neglecting the non-linear self-interactions
of the entropy
field which are negligible at sufficiently early times,
the equation of
motion for $\delta \chi$ becomes simply the equation of a
free massive field
in an unperturbed FRW metric,
\begin{eqnarray}
\ddot{\delta \chi} + 3H \dot{\delta \chi} +
\left(\frac{k^2}{a^2}+ m_\chi^2\right)\delta \chi =0\,,
\label{wave_eq_dchi_gen}
\end{eqnarray}
where $m_{\chi}^2$ has been defined in Eq.~(\ref{m_chi}).
Introducing the rescaled field $v \equiv a \delta \chi$,
and writing the wave equation in terms of conformal time
$\tau \equiv \int dt/a$, we have
\begin{eqnarray}
{v,}_{\tau\tau} + \left[k^2 - \frac{{a,}_{\tau\tau}}{a}
+ m_\chi^2 a^2\right]v=0\,,
\end{eqnarray}
where $(\ldots)_{,\tau}$
denotes the derivative with respect to $\tau$.

For the multi-field scaling solutions, we can show that
the following relations hold,
\begin{eqnarray}
&&a H = \frac{1}{(\epsilon-1)\tau},
\label{ah_multi_scaling}\\
&&\frac{{a,}_{\tau\tau}}{a}=-(\epsilon-2)a^2 H^2,\\
&&m_\chi^2 = 2 \epsilon (3-\epsilon)
H^2\,. \label{m_chi2}
\end{eqnarray}
Thus, Eq.~(\ref{wave_eq_dchi_gen}) becomes
\begin{eqnarray}
{v,}_{\tau\tau}+\left[k^2 +
\frac{\epsilon + 3\eta_\chi-2}{(\epsilon-1)^2 \tau^2}
\right]v=0\,,
\end{eqnarray}
where $\eta_\chi \equiv m_\chi^2/(3 H^2)$.

Using the usual Bunch-Davies vacuum state to normalise
the amplitude of the fluctuations on small scales,
we obtain
\begin{eqnarray}
v=\frac{\sqrt{\pi}}{2}\frac{e^{i(\nu+1/2)\frac{\pi}{2}}}
{k^{1/2}} (-k\tau)^{1/2}H_\nu^{(1)}(-k\tau)\,,
\label{v_multi_scaling}
\end{eqnarray}
where the order of the Hankel function is given by
\begin{eqnarray}
\nu^2 = \frac{9}{4}-\frac{3 \epsilon}{(\epsilon-1)^2}\,.
\label{iso_nu}
\end{eqnarray}
At late times, $-k\tau \to 0$,
making use of the asymptotic
form of the Hankel function,
\begin{eqnarray}
H_\nu^{(1)} (-k\tau) \to -i \frac{\Gamma(\nu)}{\pi}
\left(\frac{-k\tau}{2}\right)^{-\nu}\,,
\label{hankel_asym}
\end{eqnarray}
the power spectrum of $\delta \chi$ in this limit
becomes
\begin{eqnarray}
&&{\cal{P}}_{\chi}
= C_\nu^2 \frac{k^2}{a^2}(-k \tau)^{1-2\nu}\,,
\end{eqnarray}
where $C_\nu \equiv 2^{\nu-3/2} \Gamma(\nu)/\pi^{3/2}$.
Then the spectral tilt of the generated fluctuations is
\begin{eqnarray}
\Delta n_{\chi} \equiv
\frac{d \ln {\cal{P}}_{\chi}}
{d \ln k} = 3-2\nu\,,
\label{iso_tilt}
\end{eqnarray}
and  to leading order in a fast-roll expansion,
\begin{eqnarray}
 \label{tilt}
\Delta n_{\chi} \simeq \frac{2}{\epsilon}\,.
\end{eqnarray}
Thus, for a
steep exponential potential we obtain a slightly
blue spectrum, becoming scale-invariant
as $\epsilon \to \infty$.

Approximating $\nu \simeq 3/2$, on large scales we can
relate the amplitude of $ \delta \chi$ to $H$
as
\begin{eqnarray}
{{\cal{P}}}_{\chi} ^{1/2}
 = \epsilon \frac{|H|}{2\pi}\,.
\label{power_delchi}
\end{eqnarray}

\section{Isocurvature field bispectrum}
\label{app2}

In this section we calculate the intrinsic non-Gaussianity of the
field $\chi$, using the approach of Maldacena
\cite{Maldacena:2002vr}, which also includes the contribution from
sub-Hubble nonlinear interactions. Following
\cite{Maldacena:2002vr}, the three-point correlator is given by
 \begin{equation}
\langle
 \delta \chi^3 (t)\rangle
= -i \int^{t} _{-\infty } dt
\langle
[ \delta \chi^3 (t)
,H_{int} (t')]
\rangle \,, \label{Maldacena}
\end{equation}
where $H_{int}$ is the interaction Hamiltonian. Here we consider only the
cubic interaction, which is the dominant one for the three-point function,
\begin{equation}
H_{int} (t') = \int d^3 x a^3 \frac{V^{(3)}}{3!} \delta \chi^3\,,
\label{inter_hamiltonial}
\end{equation}
where $V^{(3)}$ is given in Eq.~(\ref{V3}).

Writing Eq.~(\ref{Maldacena}) in Fourier space,
\begin{eqnarray}
&&\langle \delta \chi_{{\bf k}_1}(t)
\delta \chi_{{\bf k}_2}(t)
\delta \chi_{{\bf k}_3}(t)
    \rangle = (2 \pi)^3 \delta( \sum_j {\bf k}_j) \times \nonumber\\
&&
2 Re \left( -i \delta \chi_{{\bf k}_1}(t)
\delta \chi_{{\bf k}_2}(t)
\delta \chi_{{\bf k}_3}(t)
\int_{-\infty } ^{t} dt' a^3 V^{(3)}
\delta \chi_{{\bf k}_1}^*(t')
\delta \chi_{{\bf k}_2}^*(t')
\delta \chi_{{\bf k}_3}^*(t') \right)
\,,
\end{eqnarray}
and
using the normalised free field solution
\begin{equation}
\delta \chi_{{\bf k}} = \frac{1}{a \sqrt{2k}}
e^{-ik\tau}\left( 1-\frac{i}{k\tau}\right)
\end{equation}
valid in the limit $p\ll1$,
we get
\begin{eqnarray}
&&\langle \delta \chi_{{\bf k}_1}(t)
\delta \chi_{{\bf k}_2}(t)
\delta \chi_{{\bf k}_3}(t)
    \rangle =  - (2 \pi)^3 \delta ( \sum_j {\bf k}_j)
\frac{V^{(3)}}{4 H^2}
\frac{\epsilon^2 H^4}{\prod_j k_j^3}
\times \nonumber\\
&&
 Re \left[ i
\left(1+ik_1 \tau \right)
\left(1+ik_2 \tau \right)
\left(1+ik_3 \tau \right)
e^{-i\sum_j k_j \tau} \right. \nonumber\\
&&
\tau \int_{-\infty } ^{\tau} \frac{d \tau' }{\tau'^5}
\left. \left(1-ik_1 \tau' \right)
\left(1-ik_2 \tau' \right)
\left(1-ik_3 \tau' \right)
e^{i \sum_j k_j \tau'} \right]
\,,
\end{eqnarray}
where we have taken out $V^{(3)}$ from the time integral using that
$V^{(3)}/H^2$
is constant.

Taking into account only the leading contribution in $- k_j \tau \ll 1$,
the last two lines of
the above equation yield $ \sum_j k_j^3 /4$.
Furthermore, using also Eqs.~(\ref{V3}) and (\ref{m_chi2})
to replace
$V^{(3)}$, we finally find
\begin{eqnarray}
&&\langle \delta \chi_{{\bf k}_1}(t)
\delta \chi_{{\bf k}_2}(t)
\delta \chi_{{\bf k}_3}(t)
    \rangle =  (2 \pi)^3 \delta( \sum_j {\bf k}_j)
\frac{\sum_j k_j^3}{\prod_j k_j^3}  \frac{\tilde c}{8} {\epsilon^4 H^4}
\,. \label{NGint1}
\end{eqnarray}
The intrinsic non-Gaussianity after Hubble-exit is thus of the local
form, as it is dominated by the super-Hubble evolution. Note that
using $\epsilon = 1/p$ and $H=-p/t$ we can rewrite the last term in
the equation above as
\begin{equation}
\frac{\tilde c}{8} {\epsilon^4 H^4} = \frac{\tilde c}{8 t^4}.
\end{equation}
Thus, for $\tilde c  = 1/M$ we recover the result of
\cite{Creminelli:2007aq}.
With the definition of the power spectrum for $\delta \chi$,
Eq.~(\ref{PSchi}), we can rewrite Eq.~(\ref{NGint1}) as
\begin{eqnarray}
&&\langle \delta \chi_{{\bf k}_1}(t)
\delta \chi_{{\bf k}_2}(t)
\delta \chi_{{\bf k}_3}(t)
    \rangle =  (2 \pi)^3 \delta( \sum_j {\bf k}_j)
\frac{\sum_j k_j^3}{\prod_j k_j^3}  2 \pi^4 {\tilde c} {\cal P}_\chi^2
\,, \label{NGint2}
\end{eqnarray}
which yields, using Eq.~(\ref{scalar_bispectrum_def}),
\begin{equation}
B_\chi (k_1,k_2,k_3)=
\frac{\sum_j k_j^3}{\prod_j k_j^3}
2 \pi^4 \tilde c {\cal P}_\chi ^2 \,. \label{int_bispectrum}
\end{equation}
This equation is equivalent to
Eq.~(\ref{fnl_chi_m}), confirming the result found from the large
scale nonlinear evolution in Sec.~\ref{sec5}.
Thus non-linearities on sub-Hubble scales are completely subdominant with
respect to the nonlinear classical super-Hubble evolution.

\section{Numerical results}
\label{app3} In this appendix, we check the validity of the analysis
of Secs.~\ref{sec4} and \ref{sec5} by numerically calculating the
expansion $\delta N$.

\subsection{Phase space variables}
For numerical calculations, it is more convenient to adopt phase
space variables. Thus, first we summarise the background dynamics in
terms of the phase space variables. These are defined as
\cite{Guo:2003eu,Copeland:1997et,Heard:2002dr}
\begin{eqnarray}
x_j &=& \frac{\dot{\phi}_j}{\sqrt{6} H}\,,
\label{def_xj}\\
y_j &=& \frac{\sqrt{V_j e^{-c_j \phi_j}}}{\sqrt{3} H}\,,
\label{def_yj}
\end{eqnarray}
and their evolution equations are given by
\begin{eqnarray}
\frac{d x_j}{d N} &=& -3 x_j (1- \sum_k x_k^2) - c_j
\sqrt{\frac{3}{2}}
 y_j^2\,,
\label{evol_eq_xj}\\
\frac{d y_j}{d N} &=& y_j \left(3 \sum_k x_k^2  - c_j
\sqrt{\frac{3}{2}} x_j \right)\,, \label{evol_eq_yj}
\end{eqnarray}
where $N = \log a$ and $j=1,2$ for two fields case. The Friedmann
equation gives a constraint,
\begin{equation}
\sum_j x_j^2 - \sum_j y_j^2 =1\,. \label{Fried_const}
\end{equation}
Assuming $\sum_j c_j^{-2}<1/6$, there are $4$ fixed points of the
system where $dx_j/dN = dy_j/dN =0$,
\begin{eqnarray}
A : &&\;\;\; \sum_k x_k^2 =1\,, \quad y_j =0\,.\\
B_j : &&\;\;\; x_j =\frac{c_j}{\sqrt{6}}\,, \;\;\;
y_j=-\sqrt{\frac{c_j^2}{6}-1}\,,
 \;\;\;
x_k=y_k=0\,, \;\; (\mbox{for} \;\;\; j \neq k)\,,\\
B : &&\;\;\; x_j = \frac{\sqrt{6}}{3 p} \frac{1}{c_j}\,, \;\;\; y_j
= -\sqrt{\frac{2}{c_j^2 p} \left(\frac{1}{3p}-1 \right) }\,,
\label{scaling_rel}
\end{eqnarray}
where $p=\sum_j 2/c_j^2$. A linearized stability analysis shows that
the multi-field scaling solution, $B$, always has one unstable mode.
On the other hand, the single-field-dominated scaling solutions,
$B_j$, are always stable.

In the $(x_1, x_2)$  plane, the three fixed points $B, B_1$ and
$B_2$ are connected by a straight line, which is given by
\begin{equation}
c_2 x_1 + c_1 x_2 = \frac{c_1 c_2}{\sqrt{6}}\,. \label{instability}
\end{equation}
Solutions which start close to the saddle point $B$ evolve along
this line to one of the single-field dominated solutions, $B_1$ or
$B_2$. As in the main text, we concentrate on the case in which the
background evolves from the multi-field scaling solution $B$ to
the single-field $\phi_2$-dominated scaling solution $B_2$ for
simplicity.

\subsection{Numerical scheme}

We first briefly summarise the numerical scheme. For the phase space
variables $x_j$, $y_j$ defined by Eqs.~(\ref{def_xj}) and
(\ref{def_yj}), we solve the evolution equations (\ref{evol_eq_xj})
and (\ref{evol_eq_yj}), numerically. Since we concentrate on the
case in which the background evolves from the multi-field scaling
solution, initial conditions are characterised by a point close to $B$
in the phase space. As in the main text, in the multi-field scaling
solution background, the fluctuations of the field $\varphi$ are
negligible.

Thus, for the phase space variables, we choose the initial
conditions
 \begin{eqnarray}
&& \frac{\dot{\varphi}}{\sqrt{6}H}= \frac{c_2 x_1 + c_1
x_2}{\sqrt{c_1^2 + c_2^2}}= \frac{c}{\sqrt{6}}\,,
\label{initial_dvarphi}\\
&& \frac{-\dot{\chi}}{\sqrt{6}H}= \frac{c_2 x_2 - c_1 x_1}
{\sqrt{c_1^2 + c_2^2}}\equiv z\,, \label{initial_dchi}
 \end{eqnarray}
where Eq.~(\ref{initial_dvarphi}) is given by the background scaling
solution in Eqs.~(\ref{ev_a_multi_scaling}) and
(\ref{ev_vphi_multi_scaling}). This guarantees that $(x_1, x_2)$
lies on the line given by Eq.~(\ref{instability}).

Equation~(\ref{initial_dchi}) defines the quantity $z$, which
characterises the deviation along the instability direction from
$B$. Once we fix a model, i.e. the values of $c_1$ and $c_2$, the
initial values of $x_1$ and $x_2$ are determined from
Eqs.~(\ref{initial_dvarphi}) and (\ref{initial_dchi}) for a given
initial value of $z$.

\begin{figure}[http]
 \begin{center}
\includegraphics[width=16cm]{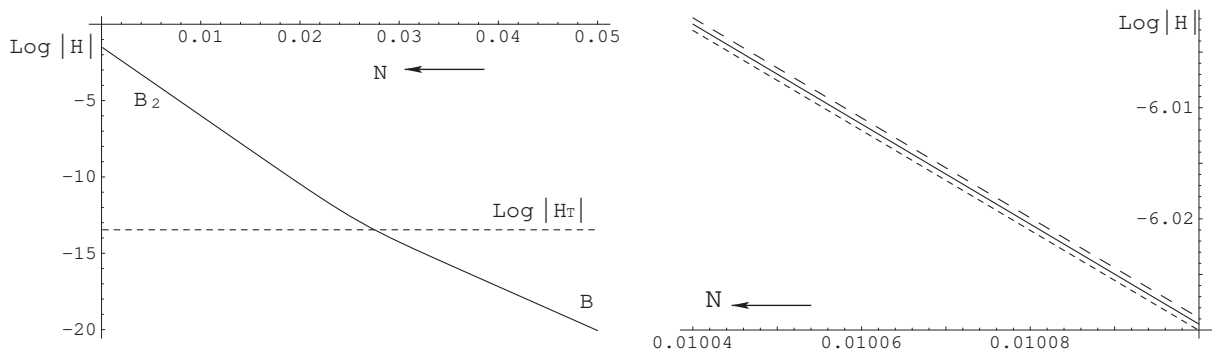}
\caption[]{Left: An example of background solution for $\log \vert H
\vert $ with initial condition $z=z_i$. We also show $\log \vert H_T
\vert $ which is determined from the amplitude of  $\delta \chi$.
Right: The same background solution shown around $N \sim 0.01005$
which is much later than the transition. We also show the solutions
with slightly different initial condition, $z=z_i + \delta z_i$,
 $z=z_i - \delta z_i$, with dashed lines.
After the transition, this difference of initial $z$ generates the
curvature perturbations which can be evaluated as the difference of
$N$ at $H=H_f$.} \label{fig_deltaN}
\end{center}
\end{figure}

For this set of $(x_1, x_2)$, $y_1$ is fixed so that 
$(x_1, x_2,y_1)$ 
is in the instability direction from $B$, and then $y_2$ is
fixed from the constraint equation (\ref{Fried_const}). Using this
set of initial values $(x_1, x_2, y_1, y_2)$, we can specify a
background trajectory by solving the evolution equation. The left
panel of Fig.~\ref{fig_deltaN} shows an example of the background
solution. For this background, we evaluate an expansion $\delta N$
from the hypersurface characterised by $H_i$ to that by $H_f$.

From Eqs.~(\ref{chi_nl}), (\ref{alphachi}) and (\ref{initial_dchi}),
we can see that a different choice of $z$ corresponds to that of
$\alpha$ and $\chi$ as
\begin{eqnarray}
z &=& \frac{\epsilon H}{\sqrt{6}} \left( 1 + \frac{1}{2} \tilde{c}
\alpha H \right) \alpha\,,
\label{rel_dz_dalpha}\\
&=& \frac {\epsilon}{\sqrt{6}} \left( 1+\frac{1}{4} \tilde{c} \chi
\right) \chi\,. \label{rel_dz_dchi}
\end{eqnarray}
Then $z_i$ is related with $\chi_i$ and it determines the Hubble
parameter at the transition $H_T$ in this scheme. In order to
calculate the derivatives of $N(z_i)$ with respect to $z_i$, we
change $z_i$ slightly and calculate the difference of the expansion
$N(z_i)$. The right panel of Fig.~\ref{fig_deltaN} shows the effect
of the change in $z_i$ on the Hubble parameter.

\subsection{Power spectrum and non-Gaussian parameter}
In the following we first check numerically the linear relation
Eq.~(\ref{powerspec}). Note that from Eqs.~(\ref{rel_dz_dalpha}) and
(\ref{rel_dz_dchi}), $\delta z \propto \delta \alpha \propto \delta
\chi$ holds at the linearised level. From
Eqs.~(\ref{rel_zeta_chi_powerspectrum}) and (\ref{PSchi}), the power
spectrum of the curvature perturbation is given by
\begin{equation}
{\cal P}_\zeta =  \barf{N_{,\chi_i}}^2 \epsilon^2 \left|
\frac{H_i}{2 \pi}  \right|^2 \,. \label{power_zeta_deltan}
\end{equation}
In our numerical scheme, from Eq.~(\ref{rel_dz_dchi}) we evaluate
$\barf{N_{,\chi_i}}$ as
\begin{equation}
\barf{N_{,\chi_i}} = \frac{\epsilon}{\sqrt{6}} \barf{N_{,z_i}} =
\frac{\epsilon}{\sqrt{6}} \frac{\barf{N}(z_i + \delta z_i) -
\barf{N}(z_i - \delta z_i)} {2 \delta z_i}\,, \label{dn_num}
\end{equation}
which is valid for sufficiently small $\delta z_i$. From
Eq.~(\ref{power_zeta_ht_2}), the power spectrum of the curvature
perturbations can also be expressed in terms of the scalar field
fluctuation $\delta \chi$ as
\begin{equation}
{\cal P}_\zeta =  \frac{\epsilon^2}{c_1^2+c_2^2} \left| \frac{H_T}{2
\pi}  \right|^2 = \frac{\epsilon^2}{c_1^2+c_2^2}
 \frac{|\delta \chi|^2_{B_2}}{|\delta \chi|^2_{B}}
\left| \frac{H_i}{2 \pi}  \right|^2 \,. \label{power_zeta_dchi}
\end{equation}
As Eq.~(\ref{power_zeta_dchi}) and Eq.~(\ref{power_zeta_deltan})
should agree, we examine whether a quantity $q$ defined by
\begin{equation}
q \equiv \frac{\barf{N_{,\chi_i}}^2  (c_1^2+ c_2^2) |\delta
\chi|^2_{B}} { |\delta \chi|^2_{B_2}}\,,
\end{equation}
becomes close enough to $1$. We find that $q=1$ is satisfied within
about 1 $\%$ accuracy in our numerical simulations.





Next, we compute numerically $f_{NL}$. From
Eq.~(\ref{rel_dz_dalpha}), we can write down $N_{,\alpha}$ and
$N_{,\alpha \alpha}$ in terms of $z$:
\begin{eqnarray}
&& N_{,\alpha} = N_{,z} \frac{d z}{d \alpha} =
\frac{\epsilon}{\sqrt{6}} H_i \left( 1+\frac{\sqrt{6}
\tilde{c}}{\epsilon} z_i \right)  N_{,z} \,, \label{rel_n_alpha_n_z}
\\
&& N_{,\alpha \alpha} = N_{,zz} \left( \frac{d z}{d \alpha}
\right)^2 + N_{,z} \frac{d^2 z}{d \alpha^2} = \frac{\epsilon^2}{6}
H_i^2 \left(1+ \frac{2 \sqrt{6} \tilde{c}} {\epsilon} z_i
\right)N_{,zz} +\frac{\epsilon \tilde{c} H_i^2}{\sqrt{6}} N_{,z}\,.
\label{rel_n_alphaalpha_n_zz}
\end{eqnarray}
In our numerical scheme, we evaluate $\barf{N}_{,z_i z_i}$ as
\begin{equation}
\barf{N}_{, z_i z_i} = \frac{\barf{N}(z_i + \delta z_i) -
2\barf{N}(z_i) + \barf{N}(z_i - \delta z_i)} {(\delta z_i)^2}\,.
\end{equation}

If the instantaneous transition approximation is valid, the
non-Gaussianity parameter $f_{NL}$ is given by
\begin{eqnarray}
f_{NL} = \frac{5}{6}\frac{N_{,\alpha \alpha}}{N_{,\alpha}^2}
\label{fnl_ito_zderiv}.
\end{eqnarray}
Thus, in order to verify the accuracy of the analytical result
$f_{NL} = -5 c_1^2/12$, we examine whether a quantity $r$ defined by
\begin{equation}
r \equiv - \frac{2 \barf{N}_{, \alpha \alpha}}{ c_1^2
\barf{N}_{,\alpha} ^2}\,,
\end{equation}
becomes close to $1$ or not.

Since we checked that the choice of $\delta z_i$ does not affect the
results for sufficiently small values, we use $\delta z_i =
10^{-5}$. It was also verified that the time evolution of $\delta
\chi$ can be described by that of the multi-field scaling solution
up to $z_i = 0.01$; thus, we take $z_i = 0.01$. Using these initial
conditions, we calculate $f_{NL}$ for various combinations of $c_1$
and $c_2$. It is worth noting that in terms of $c_1$ and $c_2$ the
fast-roll parameter is expressed as $\epsilon = c_1^2
c_2^2/2(c_1^2+c_2^2)$. For the examples we show in Table 1, the
range of the value of $\epsilon$ varies from $25$ $(c_1 =c_2= 10)$
to $400$ $(c_1 = c_2 = 40 )$. The results for $f_{NL}$ and $r$ are
summarised in Table~\ref{fnl}.

\begin{table}

\caption{\label{fnl} $c_1$ and $c_2$ dependence of the nonlinear
parameter. We also compare with the results obtained by the
instantaneous transition and the fast-roll approximations. The
deviation from $r=1$ denotes the error of these approximations. We
adopt $z_i=0.01$ and $\delta z_i = 10^{-5}$.}

 \begin{center}
\begin{tabular}{llcc}
\hline \hline $c_1$ &$c_2$ & $f_{NL}$ & $r$ \\ \hline
$10$ & $10$ & $ -40.6165 $ & $ 0.974797 $ \\
$10$ & $20$ & $ -40.9285 $ & $ 0.982283 $ \\
$10$ & $40$ & $ -41.5317 $ & $ 0.996761 $ \\
$15$ & $10$ & $ -90.4979 $ & $ 0.965311 $ \\
$15$ & $20$ & $ -92.6425 $ & $ 0.988187 $ \\
$15$ & $40$ & $ -93.5294 $ & $ 0.997647 $ \\
$20$ & $10$ & $ -161.014 $ & $ 0.966082 $ \\
$20$ & $20$ & $ -164.976 $ & $ 0.989855 $ \\
$20$ & $40$ & $ -169.365 $ & $ 1.01619 $ \\
$40$ & $10$ & $ -639.909 $ & $ 0.959865 $  \\
$40$ & $20$ & $ -659.744 $ & $ 0.989617$ \\
$40$ & $40$ & $ -664.996 $ & $ 0.997494$ \\ \hline
\end{tabular}
 \end{center}
\end{table}

From Table~\ref{fnl} we can see that for these parameters, the
results for $f_{NL}$ are accurately described by $f_{NL} = -5 c_1^2
/12$ which agrees with the result obtained analytically based on the
instantaneous transition approximation and the fast roll
approximation in the main text.
%

We can see that the analytical results deviate from the numerical
results for smaller $\epsilon$ (compare the cases with $c_1 = c_2$).
This is because the fast-roll approximation becomes worse for
smaller $\epsilon$. We can also see that, for a fixed $c_1$, the
analytical results again deviate from the numerical results for
smaller $c_2$. This is because the potential around $B$ becomes
flatter along the direction to $B_2$ for smaller $c_2$, and the
instantaneous transition approximation becomes worse.


\begin{thebibliography}{99}

\bibitem{Linde:2007fr}
  A.~Linde,
  arXiv:0705.0164 [hep-th].

\bibitem{Khoury:2001wf}
  J.~Khoury, B.~A.~Ovrut, P.~J.~Steinhardt and N.~Turok,
  Phys.\ Rev.\  D {\bf 64}, 123522 (2001)
  [arXiv:hep-th/0103239].

\bibitem{Kallosh:2001ai}
  R.~Kallosh, L.~Kofman and A.~D.~Linde,
  Phys.\ Rev.\  D {\bf 64}, 123523 (2001)
  [arXiv:hep-th/0104073].

\bibitem{Khoury:2001iy}
  J.~Khoury, B.~A.~Ovrut, P.~J.~Steinhardt and N.~Turok,
  arXiv:hep-th/0105212.

 \bibitem{Lyth:2001pf}
  D.~H.~Lyth,
  Phys.\ Lett.\  B {\bf 524}, 1 (2002)
  [arXiv:hep-ph/0106153].

\bibitem{Lyth:2003im}
  D.~H.~Lyth and D.~Wands,
  Phys.\ Rev.\  D {\bf 68}, 103515 (2003)
  [arXiv:astro-ph/0306498].


\bibitem{Creminelli:2004jg}
  P.~Creminelli, A.~Nicolis and M.~Zaldarriaga,
  Phys.\ Rev.\  D {\bf 71}, 063505 (2005)
  [arXiv:hep-th/0411270].



\bibitem{Copeland:2006tn}
  E.~J.~Copeland and D.~Wands,
  JCAP {\bf 0706}, 014 (2007)
  [arXiv:hep-th/0609183].

\bibitem{Gordon:2000hv}
  C.~Gordon, D.~Wands, B.~A.~Bassett and R.~Maartens,
  Phys.\ Rev.\  D {\bf 63}, 023506 (2001)
  [arXiv:astro-ph/0009131].


\bibitem{Notari:2002yc}
  A.~Notari and A.~Riotto,
  Nucl.\ Phys.\  B {\bf 644}, 371 (2002)
  [arXiv:hep-th/0205019].

\bibitem{Lehners:2007ac}
   J.~L.~Lehners, P.~McFadden, N.~Turok and P.~J.~Steinhardt,
  Phys.\ Rev.\  D {\bf 76}, 103501 (2007)
  [arXiv:hep-th/0702153].

\bibitem{Buchbinder:2007ad}
  E.~I.~Buchbinder, J.~Khoury and B.~A.~Ovrut,
  arXiv:hep-th/0702154.

\bibitem{Creminelli:2007aq}
  P.~Creminelli and L.~Senatore,
  arXiv:hep-th/0702165.






\bibitem{Finelli:2002we}
  F.~Finelli,
  Phys.\ Lett.\  B {\bf 545}, 1 (2002)
  [arXiv:hep-th/0206112].

\bibitem{Guo:2003eu}
  Z.~K.~Guo, Y.~S.~Piao and Y.~Z.~Zhang,
  Phys.\ Lett.\  B {\bf 568}, 1 (2003)
  [arXiv:hep-th/0304048].

\bibitem{KW}
   K.~Koyama and D.~Wands,
  JCAP {\bf 0704}, 008 (2007)
  [arXiv:hep-th/0703040].

\bibitem{Buchbinder:2007tw}
  E.~I.~Buchbinder, J.~Khoury and B.~A.~Ovrut,
  arXiv:0706.3903 [hep-th].


\bibitem{KMW}
 K.~Koyama, S.~Mizuno and D.~Wands,
  Class.\ Quant.\ Grav.\  {\bf 24}, 3919 (2007)
  [arXiv:0704.1152 [hep-th]].

\bibitem{Tolley:2007nq}
  A.~J.~Tolley and D.~H.~Wesley,
  JCAP {\bf 0705}, 006 (2007)
  [arXiv:hep-th/0703101].

\bibitem{Bartolo:2004if}
  N.~Bartolo, E.~Komatsu, S.~Matarrese and A.~Riotto,
  Phys.\ Rept.\  {\bf 402}, 103 (2004)
  [arXiv:astro-ph/0406398].

\bibitem{Planck}
http://www.rssd.esa.int/index.php?project=Planck

\bibitem{Maldacena:2002vr}
  J.~M.~Maldacena,
  JHEP {\bf 0305}, 013 (2003)
  [arXiv:astro-ph/0210603].


\bibitem{Malik:1998gy}
  K.~A.~Malik and D.~Wands,
  Phys.\ Rev.\  D {\bf 59}, 123501 (1999)
  [arXiv:astro-ph/9812204].

\bibitem{Starobinsky:1986fx}
  A.~A.~Starobinsky,
  JETP Lett.\  {\bf 42}, 152 (1985)
  [Pisma Zh.\ Eksp.\ Teor.\ Fiz.\  {\bf 42}, 124 (1985)].

\bibitem{Sasaki:1995aw}
  M.~Sasaki and E.~D.~Stewart,
  Prog.\ Theor.\ Phys.\  {\bf 95}, 71 (1996)
  [arXiv:astro-ph/9507001].

\bibitem{Sasaki:1998ug}
  M.~Sasaki and T.~Tanaka,
  Prog.\ Theor.\ Phys.\  {\bf 99}, 763 (1998)
  [arXiv:gr-qc/9801017].


\bibitem{Vernizzi:2006ve}
  F.~Vernizzi and D.~Wands,
  JCAP {\bf 0605} (2006) 019
  [arXiv:astro-ph/0603799].



\bibitem{Lyth:2005fi}
  D.~H.~Lyth and Y.~Rodriguez,
  Phys.\ Rev.\ Lett.\  {\bf 95}, 121302 (2005)
  [arXiv:astro-ph/0504045].







\bibitem{Byrnes:2007tm}
  C.~T.~Byrnes, K.~Koyama, M.~Sasaki and D.~Wands,
  arXiv:0705.4096 [hep-th].

\bibitem{Salopek:1990jq}
  D.~S.~Salopek and J.~R.~Bond,
  Phys.\ Rev.\  D {\bf 42}, 3936 (1990).




\bibitem{Wands:2000dp}
  D.~Wands, K.~A.~Malik, D.~H.~Lyth and A.~R.~Liddle,
  Phys.\ Rev.\  D {\bf 62}, 043527 (2000)
  [arXiv:astro-ph/0003278].

\bibitem{Seery:2005gb}
  D.~Seery and J.~E.~Lidsey,
  JCAP {\bf 0509}, 011 (2005)
  [arXiv:astro-ph/0506056].

\bibitem{Komatsu:2008hk}
  E.~Komatsu {\it et al.}  [WMAP Collaboration],
  Astrophys.\ J.\ Suppl.\  {\bf 180}, 330 (2009)
  [arXiv:0803.0547 [astro-ph]].

\bibitem{Smith:2009jr}
  K.~M.~Smith, L.~Senatore and M.~Zaldarriaga,
  arXiv:0901.2572 [astro-ph].




\bibitem{Creminelli:2006xe}
  P.~Creminelli, M.~A.~Luty, A.~Nicolis and L.~Senatore,
  JHEP {\bf 0612}, 080 (2006)
  [arXiv:hep-th/0606090].


\bibitem{Copeland:1997et}
  E.~J.~Copeland, A.~R.~Liddle and D.~Wands,
  Phys.\ Rev.\  D {\bf 57}, 4686 (1998)
  [arXiv:gr-qc/9711068].


\bibitem{Heard:2002dr}
  I.~P.~C.~Heard and D.~Wands,
  Class.\ Quant.\ Grav.\  {\bf 19}, 5435 (2002)
  [arXiv:gr-qc/0206085].



















\end{thebibliography}
\end{document}